%% file: SGRB_proceeding_rowlinson.tex
\documentclass[12pt]{article}
\usepackage{graphicx}

\newcommand\pubnumber{Article 22 in eConf C1304143}
\newcommand\pubdate{\today}

\def\napoli{Anton Pannekoek Institute\\
Universiteit van Amsterdam, Science Park, Amsterdam}

\def\Title#1{\begin{center} {\Large #1 } \end{center}}
\def\Author#1{\begin{center}{ \sc #1} \end{center}}
\def\Address#1{\begin{center}{ \it #1} \end{center}}

\newcommand\pubblock{\rightline{\begin{tabular}{l} \pubnumber\\
         \pubdate  \end{tabular}}}
\newenvironment{Abstract}{\begin{quotation}  }{\end{quotation}}
\newenvironment{Presented}{\begin{quotation} \begin{center} 
             PRESENTED AT\end{center}\bigskip 
      \begin{center}\begin{large}}{\end{large}\end{center} \end{quotation}}
\def\Acknowledgements{\bigskip  \bigskip \begin{center} \begin{large}
             \bf ACKNOWLEDGEMENTS \end{large}\end{center}}

\input econfmacros.tex

\begin{document}
\begin{titlepage}
\pubblock

\vfill
\Title{Studying the multi-wavelength signals from short GRBs}
\vfill
\Author{ A. Rowlinson}
\Address{\napoli}
\vfill
\begin{Abstract}
Since the first host galaxies and afterglows of short GRBs were identified, they have remained very difficult to study: their multiwavelenth afterglows are notoriously faint and host galaxy identification often relies upon minimalising a chance alignment probability. Despite these observational challenges, there is now a sufficiently large sample to constrain the properties of the wider population and, in this review talk, I will summarise the current multi-wavelength observations of short GRBs. Additionally, I will describe how these observed data are able to both support and challenge the standard theoretical models of the progenitors and central engines. Looking towards the future, due to technological and theoretical advances, we are about to enter an exciting era for the study of short GRBs. We will be able to search for predicted counterparts in wide-field multi-wavelength transient searches and have the tantalising prospect of finding the very first ``smoking gun'' signal from the progenitor via the detection of gravitational waves.
\end{Abstract}
\vfill
\begin{Presented}
Huntsville Gamma Ray Burst Symposium\\
Nashville, Tennessee, 14-18 April 2013 
\end{Presented}
\vfill
\end{titlepage}
\def\thefootnote{\fnsymbol{footnote}}
\setcounter{footnote}{0}

\section{Introduction}

From the first results from the Burst and Transient Source Experiment (BATSE) \cite{fishman1985}, it was clear that there is a bimodal distribution in the observed durations of Gamma-Ray Bursts (GRBs), with short GRBs (SGRBs) having a duration of less than $\sim$2 s \cite{kouveliotou1993}. The brief flash of gamma-rays remained the only observational clue about these events until the launch of the {\it Swift} Satellite in November 2004 \cite{gehrels2004}. {\it Swift} has revolutionised the study of SGRBs, as the rapid slewing capability enabled the routine localisation of the rapidly fading X-ray afterglows. With the rapid response of optical observers, the extremely faint optical afterglows were identified along with the first host galaxies \cite{gehrels2005, fox2005, hjorth2005}. The initial classification of SGRBs as a different population from the Long GRBs (LGRBs), based on their durations, gained further supporting evidence as their multiwavelength properties such as their afterglows, locations in host galaxies and host galaxy properties were also found to be different to the LGRB population.

The LGRB population are now believed to originate from the collapse of a rapidly rotating Wolf-Rayet star (e.g. \cite{woosley1993}). Since GRB 980425 and the detection of SN 1998bw \cite{galama1998, kulkarni1998}, several LGRBs have now been associated with Ibc supernovae providing vital supporting evidence for this theory. However, to date, there has been no such "smoking gun" signal for SGRBs. The most highly supported progenitor model for SGRBs is the merger of two compact objects, typically a neutron star and a black hole or two neutron stars \cite{lattimer1976, eichler1989}. The outcome of this merger is still debated, but is predicted to be a black hole \cite{lattimer1976, eichler1989} or a rapidly rotating hypermassive neutron star with extreme magnetic fields (magnetar) \cite{dai1998a, dai2006}.

In this conference proceeding, I summarise recent progress in studying the host galaxies and multi-wavelength afterglows of SGRBs and how these data have been used to constrain the properties of their progenitors. Section 2 focuses on the prompt emission properties of SGRBs. Their positions relative to host galaxies and host galaxy properties are reviewed in Section 3. The multi-wavelength afterglows are presented in Section 4, highlighting recent developments showing evidence of energy injection which may be associated with a magnetar central engine. Finally Section 5 looks forward to future progress in understanding SGRBs.

\section{How are SGRBs initially identified?}

In addition to the identification of a bimodal classification of GRBs, SGRBs were also found to be spectrally harder than LGRBs on average \cite{kouveliotou1993} and, unlike LGRBs, have negligible lag times between the hard and soft emission \cite{norris2006}. However, from the beginning, it was clear that there is a significant overlap between the two catagories making it very difficult to unambiguously classify some GRBs. This overlap is also significantly dependent upon the properties of the instrument used to detect them \cite{qin2013, bromberg2013}.

Further confusion occured when a new class of GRBs was identified. GRB 060614 was a LGRB (T$_{90}\sim$100 s) with no associated supernova to deep limits and other properties more consistent with the SGRB population (e.g. \cite{gehrels2006, fynbo2006,dellavalle2006}). The prompt emission comprised of a short initial pulse, which looked like a typical SGRB, followed by a softer long pulse of extended emission. There is now a much larger sample of these unusual GRBs, known as SGRBs with extended emission. They typically have brighter and longer lived afterglows than SGRBs \cite{norris2011}. Additionally, some SGRBs (but not all) have been shown to have extended emission \cite{norris2010}. They are generally thought to share the same progenitor as SGRBs, however this has not been proven to date so caution is required when including them in samples of SGRBs.

This section has shown that using the prompt emission alone does not enable an unambiguous classification of SGRBs. Although there are clearly two distributions of GRBs, we need to use other multiwavelength properties to disentangle individual SGRBs from the LGRB population.

\section{Where do SGRBs come from?}

In direct contrast to the sample of LGRBs the host galaxies of SGRBs are highly variable, ranging from actively star forming galaxies to elliptical galaxies with old stellar populations. The host galaxies of SGRBs most closely resemble typical field galaxies \cite{berger2009} and the majority are late-type galaxies \cite{fong2010}.

The majority of SGRBs also occur significantly offset from their host galaxies (e.g. \cite{fong2010, church2011}). This offset is consistent with the progenitor theory, neutron stars and black holes receive a kick at their formation that can reach give the binary system velocities sufficiently high for them to escape their host galaxy. Given the long merger timescales expected and high velocities, they can travel large distances from their host galaxies \cite{tutukov1994}.

By studying the host galaxies in detail, we can place further constraints on the progenitor systems. GRB 0800905A, with the lowest confirmed redshift for a SGRB and a clearly resolved host galaxy, enabled the first detailed spatially resolved spectroscopy of a SGRB host galaxy \cite{rowlinson2010a}. The SGRB occurred offset from the northern spiral arm of its host galaxy (as shown in Figure 1). The northern spiral arm has a higher metallicity, lower star formation rate and older population of stars in comparison to the southern spiral arm. This result is significantly different to the environments of LGRBs and is consistent with the compact binary merger progenitor theory.

\begin{figure}[htb]
\centering
\includegraphics[height=2.7in]{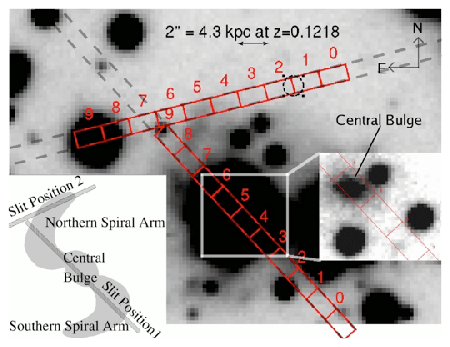}
\includegraphics[height=2.7in]{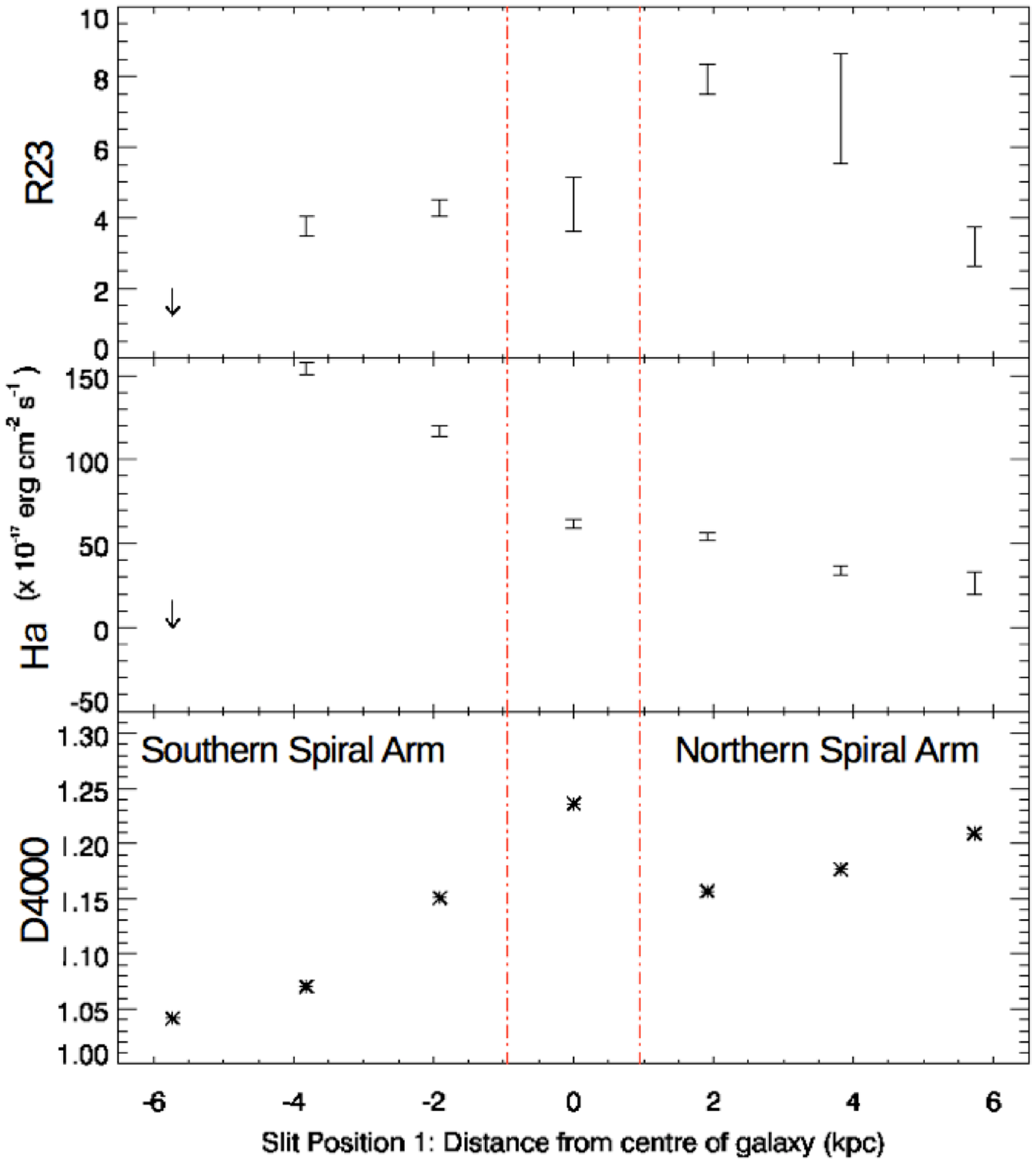}
\caption{Left: The host galaxy of GRB 080905A (sketched in the inset on the bottom left), the GRB location is marked with the dashed circle and the red boxes mark the positions of the slits (image from \cite{rowlinson2010a}). Inset on the bottom right is a K-band observation of the central bulge of the host. Right: The results of spectroscopy of GRB 080905A with distance from the centre of the host. The top row shows the R23 metallicity, the middle row shows the H$_{\alpha}$ flux (a tracer for star formation) and the bottom row shows the D4000 index (larger D4000 values indicate older stellar populations \cite{bruzual1983})}
\label{fig:080905A}
\end{figure}

However, the offsets of SGRBs often make it very difficult to identify their host galaxies (e.g. \cite{levan2007}). This host ambiguity has led to the identification of a class of SGRBs without an identified host galaxy, known as hostless SGRBs. Often there are several candidate host galaxies and a statistical argument is typically used to identify the counterpart. The probability of chance alignment of the GRB and the host galaxy is calculated using the sky density of galaxies (e.g. \cite{bloom2007}). GRB 090515 is a classic example of a hostless SGRB \cite{rowlinson2010b} as there are a number of candidate hosts in the field but it is not clear which is the actual host galaxy. GRB 090515 can be associated with a host galaxy at z$\sim$0.4, however there is a second candidate (at z$\sim$0.6) with only a slightly higher probability of chance alignment \cite{berger2010}. While it can be shown statistically that SGRBs lie at relatively low redshift \cite{tunnicliffe} and that most of the host associations are likely to be correct, it remains difficult to unambiguously identify the host galaxy of many SGRBs. Hence, caution is required when using the restframe properties of SGRBs, as almost all SGRB redshifts are from their host galaxy associations.

\section{What can be learnt from the SGRB afterglows?}

The multiwavelength afterglows of SGRBs have been observed from X-ray to, on rare occasions, radio frequencies. In this section, I focus on the properties of X-ray and optical afterglows in comparison to theoretical models.

The X-ray afterglows of SGRBs show a wide range of behaviour, some have flares and plateaus whilst others are faint and fade rapidly \cite{rowlinson2013}. The X-ray afterglows of SGRBs have been found to be fainter at 11 hours after the trigger time than the LGRB sample \cite{nysewander2009}. Additionally, although SGRB flares share many properties with their LGRB counterparts, they have significantly lower luminosities \cite{margutti2011}. A correlation between the luminosity and duration of plateaus observed in LGRB lightcurves has been observed \cite{dainotti2010} and, by fitting the BAT-XRT lightcurves of SGRBs, SGRB plateaus have also been identified and lie upon the same correlation, though at shorter durations and higher luminosities \cite{rowlinson2013}. The plateaus in LGRB lightcurves are typically associated with prolonged energy injection. However, the typical progenitor models of SGRBs predict that all accretion onto the black hole occurs within the first 2 seconds (e.g. \cite{rezzolla2011}), with a small amount of material on highly eccentric orbits which could accrete at late times \cite{rosswog2007}, this means that prolonged energy injection is not expected. Therefore, SGRB X-ray afterglows appear to be comparable to those of LGRBs, only evolving faster and more rapidly fading, and show also evidence of prolonged energy injection.

SGRBs have very faint optical afterglows, significantly fainter than those of LGRBs \cite{nysewander2009, kann2011}, making rapid follow-up with large optical telescopes vital. For instance GRB 090515 had the faintest optical afterglow detected to date with r$\sim$26.4, at 2 hours after the GRB \cite{rowlinson2010b}. When extrapolating the early time optical and X-ray observations to 1 keV, it becomes apparent that the optical afterglows are often very faint compared to their own X-ray observations, for example GRB 090515 is shown in Figure 2. This difference could potentially be explained using absorption, however it is possible to predict the expected optical absorption using the N$_H$ absorption detected in the X-ray spectrum. In many cases the predicted absorption is significantly lower than that required to explain the difference between the optical and X-ray observations. This suggests that the X-ray flux may be originating from a different emission mechanism than the optical flux \cite{rowlinson2013}.

\begin{figure}[htb]
\centering
\includegraphics[height=3.2in]{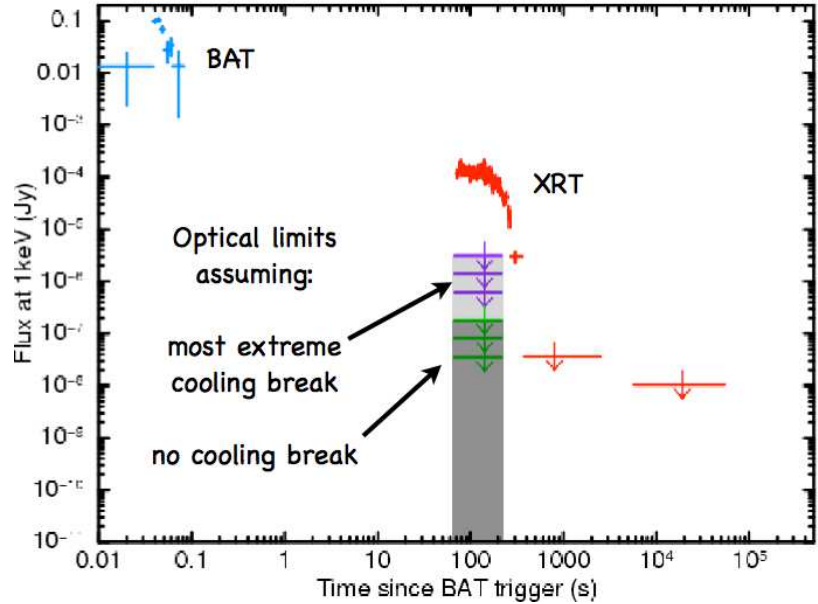}
\caption{The lightcurve of GRB 090515 extrapolated to 1 keV. The optical observations were extrapolated using the fitted spectral index with an extreme cooling break between the X-ray and optical and with no cooling break (for further details see \cite{rowlinson2013}). If the X-ray and optical emission were consistent with each other, the XRT data points would pass through the shaded region.}
\label{fig:magnet}
\end{figure}

Therefore, in SGRBs we have observed X-ray plateau emission which is inconsistent with the optical observations and the black hole central engine model. However, a black hole is not the only predicted outcome of a binary neutron star merger as a magnetar could also be formed \cite{dai1998a, dai2006}. With recent observations of neutron stars with masses of $\sim$2 M$_{\odot}$ \cite{demorest2010, antoniadis2013}, forming a hypermassive neutron star (magnetar) is a realistic potential outcome of the merger of two neutron stars \cite{ozel2010}. Recent constraints on the neutron star equation of state, resulting from these massive neutron stars, have been applied to simulations of neutron star mergers and these simulations predict the formation of a magnetar \cite{hotokezaka2013, giacomazzo2013}. The newly formed magnetar is expected to be spinning at high velocities and will be spinning down very fast via the emission of gravitational waves (at early times) and electromagnetic dipolar radiation \cite{zhang2001}. This model predicts a plateau in the lightcurve followed by a shallow decay as the magnetar spins down. When the magnetar is first formed, its rapid rotation can support a very large mass ($>1.5 \times M_{max}$, where $M_{max}$ is the maximum possible stable mass of a neutron star) but, if the magnetar is more massive than the maximum stable mass of a neutron star, it can spin down to a point where it can no longer support itself via rotation and the magnetar will collapse to form a black hole. This immediately shuts off the energy injection, causing a rapid decay in the lightcurve. The possible outcomes of this model are illustrated in Figure 3 and two example GRBs fitted with this model are shown in Figure 4.

\begin{figure}[htb]
\centering
\includegraphics[height=2.6in]{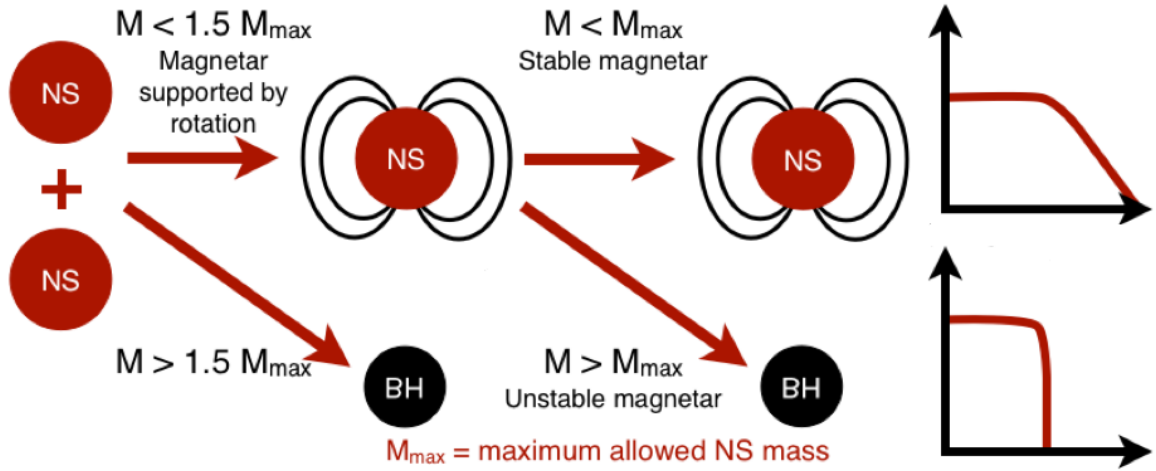}
\caption{This diagram represents the three different outcomes of the merger of a binary neutron star (NS) merger. The outcome is dependent upon the mass (M) of the central object formed and the maximum possible mass of a neutron star (M$_{max}$). On the right are sketches of the expected lightcurves if a stable (top) or an unstable magnetar (bottom) is formed.}
\label{fig:magnet}
\end{figure}

\begin{figure}[htb]
\centering
\includegraphics[height=2.95in]{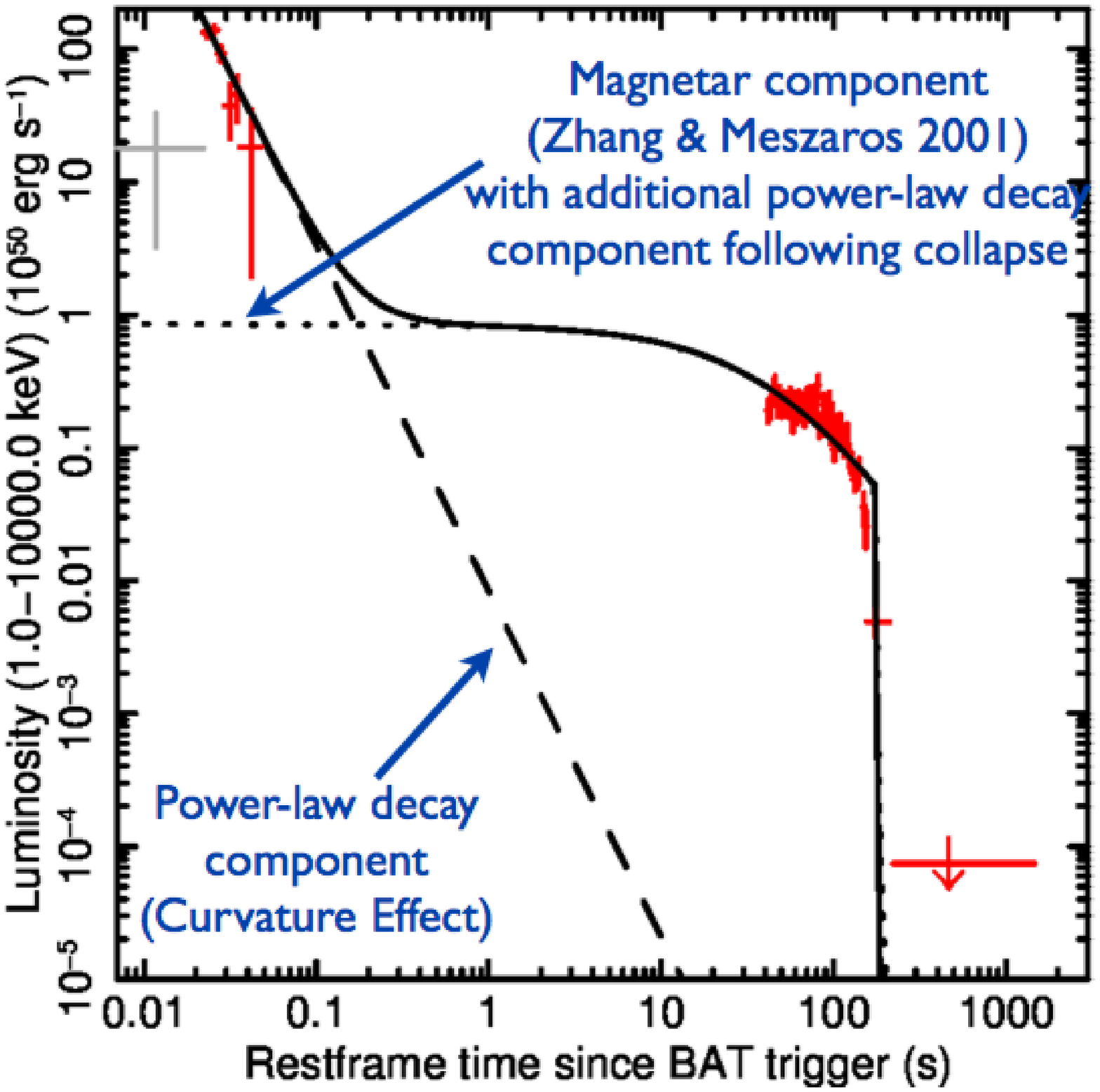}
\includegraphics[height=2.95in]{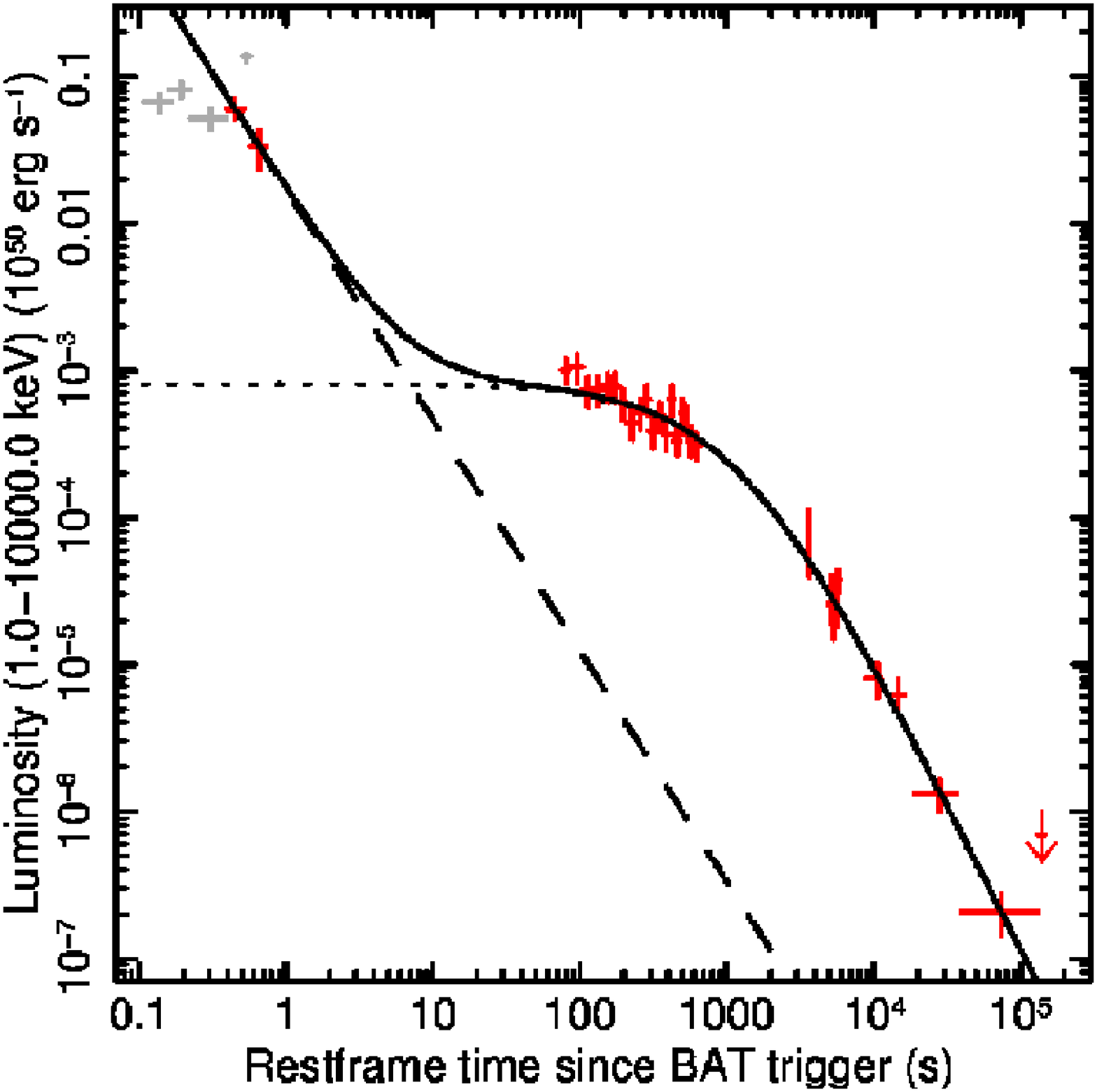}
\caption{Left: The lightcurve of GRB 090515 is an example of an unstable magnetar central engine. Right: GRB 061201 is an stable magnetar example. In both restframe BAT-XRT lightcurves, the underlying power-law decay is illustrated with a dashed line and the dotted line represents the magnetar component.}
\label{fig:magnet}
\end{figure}

All {\it Swift} SGRBs (with T$_{90} \le$2 s and sufficient data) have been fitted with the magnetar model and a large number were found to be consistent with having energy injection from a magnetar central engine \cite{rowlinson2013}. The magnetic field and spin period solutions for the SGRB sample are shown in Figure 5 in comparison to a sample of LGRBs which are also thought to have magnetar central engines. The full analysis (described in \cite{rowlinson2013}) required assumptions about beaming, efficiency and redshift. However, while these assumptions affect the actual fitted values of the magnetic field and spin period of the magnetars, they do not affect the applicability of the magnetar central engine model. Note, this magnetar central engine model may also hold the potential of explaining the extended emission observed in some SGRBs \cite{metzger2008, gompertz2013}.

\begin{figure}[htb]
\centering
\includegraphics[height=2.95in]{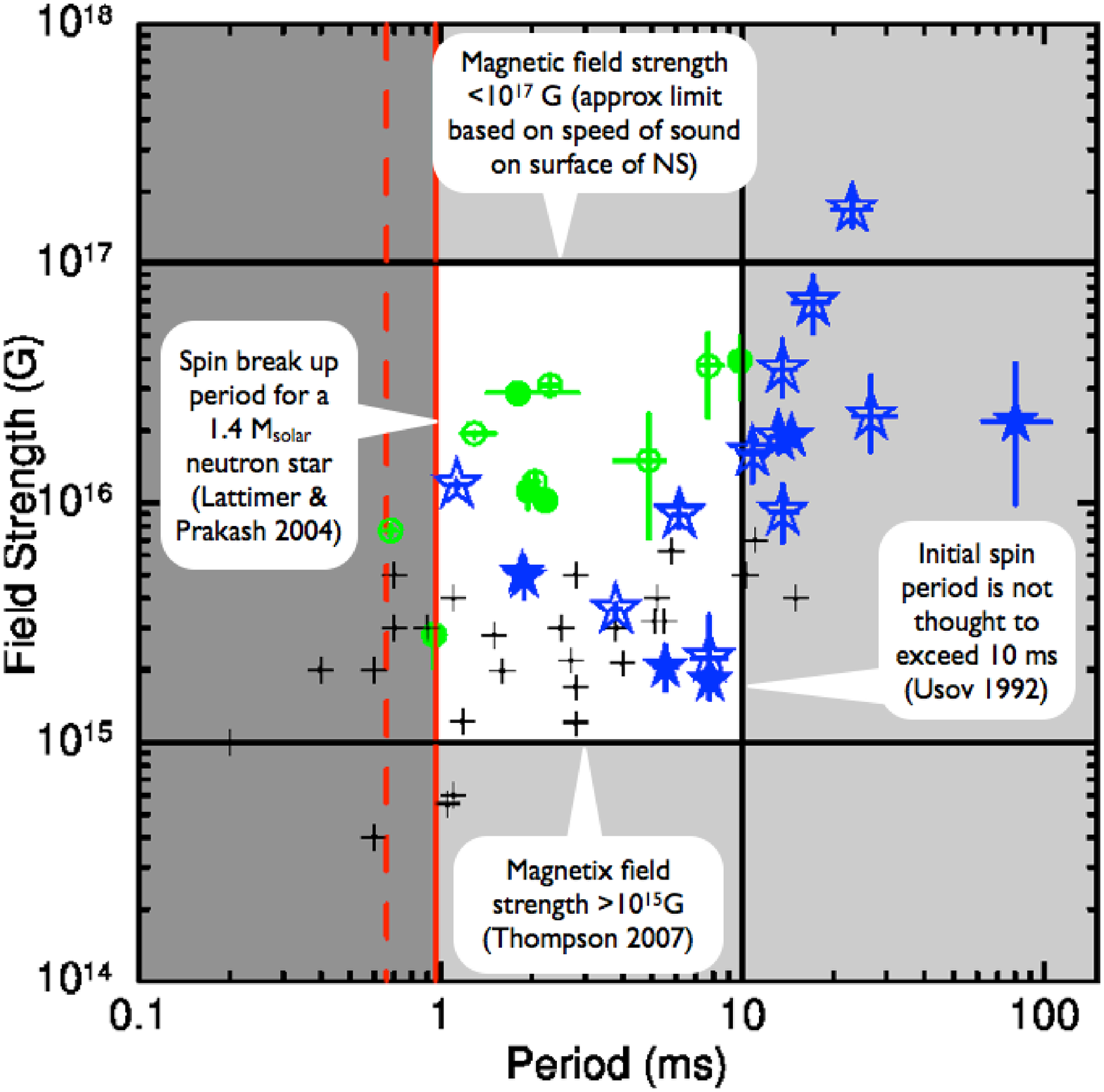}
\includegraphics[height=2.95in]{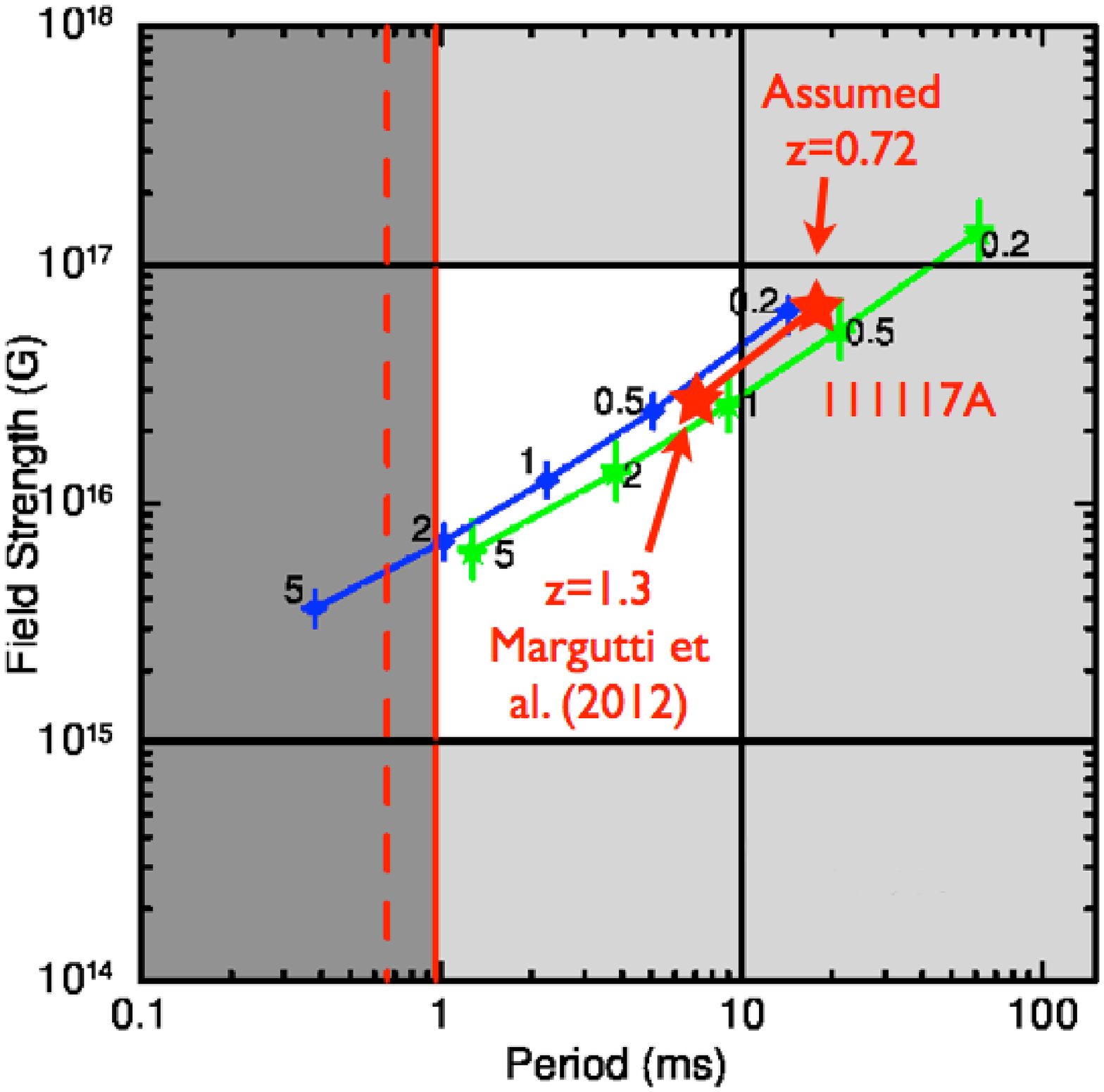}
\caption{Left: The spin periods and magnetic field strengths of the SGRB sample plotted with the LGRB candidates \cite{lyons2009,dallosso2011,bernardini2012}, many lie within the white region as expected. The blue stars are the stable magnetar candidates and the green cricles are the unstable candidates. Right: Many of the SGRBs in the sample do not have redshifts so an average redshift was assumed, the green and blue lines represent the different redshift solutions for GRBs 080702A and 060801 respectively. Recently a redshift was published for GRB 111117A \cite{margutti2012}, the red line shows that the magnetar solutions for this GRB have moved from the shaded region to the expected region when using the published redshift.}
\label{fig:magnet}
\end{figure}

\section{What does the future hold for SGRBs?}

The inspiral and merger of the compact binary systems thought to produce SGRBs are also a predicted standard candle for the observation of gravitational waves. To date, gravitational wave observatories have been able to place constraining limits on 2 SGRBs which were observationally close to nearby galaxies (GRB 070201A \cite{abbott2008} and GRB 051103 \cite{abadie2012}) however the majority of SGRBs lie far beyond the detection limits. Advanced-LIGO (Laser Interferometer Gravitational-wave Observatory), expected online from 2015, will be able to start probing out to sufficiently large distances to start placing interesting constraints on the progenitors of SGRBs \cite{aasi2013}. Additionally, if a SGRB is sufficiently nearby, the gravitational wave observations hold the potential of answering the question of what the central engines of SGRBs are, as newly formed magnetars are predicted to emit gravitational waves \cite{zhang2001,corsi2009}.

Alongside the implementation of Advanced-LIGO, there is increased interest in the additional isotropic emission mechanisms that may be associated with the merger of a compact binary system. During the binary merger, a small amount of material will be ejected at high velocity. This material will interact with itself, creating radioactive elements which emit as they undergo radioactive decay, producing an isotropic optical signal referred to as a macronova or a kilonova \cite{li1998, kulkarni2005, metzger2010}. This material will also interact with the surrounding inter-stellar medium giving a radio flare \cite{piran2013}. Additionally, if the system is powered by a magnetar, there may be a strong proto-magnetar wind that will inject additional energy into the surrounding environment  \cite{zhang2013}. The proto-magnetar wind may also be travelling at relativistic velocities providing significantly brighter signals on different timescales and at all observing frequencies \cite{gao2013}. Many widefield optical and radio observatories are now starting to come online which may be able to detect these predicted signals.

\section{Conclusions}

This conference proceeding has highlighted many open issues with the study of SGRBs including:

\begin{itemize}
    \item the difficulty in identifing SGRBs using the prompt emission alone due to overlapping distributions and the group of SGRBs with extended emission.
    \item host galaxies typically rely on chance alignment calculations due to significant offsets and can be ambiguous.
    \item the vast majority of SGRB redshifts are from their associated host galaxies and rely upon the correct host identification.
    \item their afterglows fade very rapidly, making them more difficult to detect and lead to sparsely sampled lightcurves.
\end{itemize}

Despite these caveats, we now have a sample of well-studied SGRBs that can be used to place interesting constraints on their progenitor systems and central engines. The multiwavelength observations are consistent with the theorised compact binary merger model however this progenitor remains unconfirmed.

The X-ray afterglows of SGRBs have been shown to have evidence of energy injection at early times, which is inconsistent with the typical black hole central engine model. However, the observed energy injection can be explained if the central engine is instead a newly formed magnetar.

With the new and upcoming observatories, now is the time for significant progress in the study of SGRBs and holds the potential of many exciting discoveries. However, in order to reap the full rewards, speed is the issue; SGRBs require good localisations and rapid follow-up with deep, multiwavelength observations. 

\Acknowledgements

I thank the conference organisers for inviting me to give this talk. Additionally, I thank Paul O'Brien and Nial Tanvir for their support throughout my PhD and beyond.

\end{document}

%% file: econfmacros.tex
\def\beq{\begin{equation}}
\def\eeq#1{\label{#1}\end{equation}}
\def\eeqn{\end{equation}}

\def\beqa{\begin{eqnarray}}
\def\eeqa#1{\label{#1}\end{eqnarray}}
\def\eeqan{\end{eqnarray}}

\let\bar=\overbar

\def\Dslash{\not{\hbox{\kern-4pt $D$}}}
\def\dslash{\not{\hbox{\kern-2pt $\del$}}}

\def\msb{{\bar{\ssstyle M \kern -1pt S}}}




%% file: SGRB_proceeding_rowlinson.bbl
\begin{thebibliography}{99}

\bibitem{fishman1985} Fishman G.~J., Meegan C.~A., Parnell T.~A., Wilson R.~B., Paciesas W., Mateson J.~L., Cline T.~L., Teegarden B.~J., 1985, ICRC, 3, 343 
\bibitem{kouveliotou1993} Kouveliotou C., Meegan C.~A., Fishman G.~J., Bhat N.~P., Briggs M.~S., Koshut T.~M., Paciesas W.~S., Pendleton G.~N., 1993, ApJ, 413, L101 
\bibitem{gehrels2004} Gehrels N., et al., 2004, ApJ, 611, 1005 
\bibitem{gehrels2005} Gehrels N., et al., 2005, Nature, 437, 851 
\bibitem{fox2005} Fox D.~B., et al., 2005, Nature, 437, 845 
\bibitem{hjorth2005} Hjorth J., et al., 2005, Nature, 437, 859 
\bibitem{woosley1993} Woosley S.~E., 1993, ApJ, 405, 273 
\bibitem{galama1998} Galama T.~J., et al., 1998, Nature, 395, 670 
\bibitem{kulkarni1998} Kulkarni S.~R., et al., 1998, Nature, 395, 663 
\bibitem{lattimer1976} Lattimer, J.~M., Schramm, D.~N., 1976, ApJ, 210, 549 
\bibitem{eichler1989} Eichler, D., Livio, M., Piran, T., Schramm, D.~N., 1989, Nature, 340, 126 
\bibitem{dai1998a} Dai Z.~G., Lu T., 1998a, A\&A, 333, L87 
\bibitem{dai2006} Dai Z.~G., Wang X.~Y., Wu X.~F., Zhang B., 2006, Sci, 311, 1127 
\bibitem{norris2006} Norris J.~P., Bonnell J.~T., 2006, ApJ, 643, 266 
\bibitem{qin2013} Qin Y., et al., 2013, ApJ, 763, 15 
\bibitem{bromberg2013} Bromberg O., Nakar E., Piran T., Sari R., 2013, ApJ, 764, 179 
\bibitem{gehrels2006} Gehrels N., et al., 2006, Nature, 444, 1044 
\bibitem{fynbo2006} Fynbo J.~P.~U., et al., 2006, Nature, 444, 1047 
\bibitem{dellavalle2006} Della Valle M., et al., 2006, Nature, 444, 1050 
\bibitem{norris2011} Norris J.~P., Gehrels N., Scargle J.~D., 2011, ApJ, 735, 23 
\bibitem{norris2010} Norris J.~P., Gehrels N., Scargle J.~D., 2010, ApJ, 717, 411 
\bibitem{berger2009} Berger E., 2009, ApJ, 690, 231 
\bibitem{fong2010} Fong W., Berger E., Fox D.~B., 2010, ApJ, 708, 9 
\bibitem{church2011} Church R.~P., Levan A.~J., Davies M.~B., Tanvir N., 2011, MNRAS, 413, 2004 
\bibitem{tutukov1994} Tutukov A.~V., Yungelson L.~R., 1994, MNRAS, 268, 871 
\bibitem{rowlinson2010a} Rowlinson A., et al., 2010a, MNRAS, 408, 383
\bibitem{bruzual1983} Bruzual A.~G., 1983, ApJ, 273, 105
\bibitem{levan2007} Levan A.~J., et al., 2007, MNRAS, 378, 1439 
\bibitem{bloom2007} Bloom J.~S., et al., 2007, ApJ, 654, 878 
\bibitem{rowlinson2010b} Rowlinson, A., et al., 2010b, MNRAS, 409, 531
\bibitem{berger2010} Berger E., 2010, ApJ, 722, 1946 
\bibitem{tunnicliffe} Tunnicliffe, R., et al., MNRAS Submitted
\bibitem{rowlinson2013} Rowlinson A., O'Brien P.~T., Metzger B.~D., Tanvir N.~R., Levan A.~J., 2013, MNRAS, 430, 1061 
\bibitem{nysewander2009} Nysewander M., Fruchter A.~S., Pe'er A., 2009, ApJ, 701, 824 
\bibitem{margutti2011} Margutti R., et al., 2011, MNRAS, 417, 2144 
\bibitem{dainotti2010} Dainotti M.~G., Willingale R., Capozziello S., Fabrizio Cardone V., Ostrowski M., 2010, ApJ, 722, L215 
\bibitem{rezzolla2011} Rezzolla L., Giacomazzo B., Baiotti L., Granot J., Kouveliotou C., Aloy M.~A., 2011, ApJ, 732, L6 
\bibitem{rosswog2007} Rosswog S., 2007, MNRAS, 376, L48 
\bibitem{kann2011} Kann D.~A., et al., 2011, ApJ, 734, 96 
\bibitem{demorest2010} Demorest P.~B., Pennucci T., Ransom S.~M., Roberts M.~S.~E., Hessels J.~W.~T., 2010, Nature, 467, 1081 
\bibitem{antoniadis2013} Antoniadis J., et al., 2013, Sci, 340, 448 
\bibitem{ozel2010} {\"O}zel F., Psaltis D., Ransom S., Demorest P., Alford M., 2010, ApJ, 724, L199 
\bibitem{hotokezaka2013} Hotokezaka K., Kyutoku K., Shibata M., 2013, PhRvD, 87, 044001 
\bibitem{giacomazzo2013} Giacomazzo B., Perna R., 2013, ApJ, 771, L26 
\bibitem{zhang2001} Zhang B., M{\'e}sz{\'a}ros P., 2001, ApJ, 552, L35 
\bibitem{metzger2008} Metzger B.~D., Quataert E., Thompson T.~A., 2008, MNRAS, 385, 1455 
\bibitem{gompertz2013} Gompertz B.~P., O'Brien P.~T., Wynn G.~A., Rowlinson A., 2013, MNRAS, 431, 1745 
\bibitem{lyons2009} Lyons N., O'Brien P.~T., Zhang B., Willingale R., Troja E., Starling R.~L.~C., 2010, MNRAS, 402, 705 
\bibitem{dallosso2011} Dall'Osso S., Stratta G., Guetta D., Covino S., de Cesare G., Stella L., 2011, A\&A, 526, A121 
\bibitem{bernardini2012} Bernardini M.~G., Margutti R., Mao J., Zaninoni E., Chincarini G., 2012, A\&A, 539, A3 
\bibitem{margutti2012} Margutti R., et al., 2012, ApJ, 756, 63
\bibitem{lattimer2004} Lattimer J.~M., Prakash M., 2004, Sci, 304, 536
\bibitem{usov1992} Usov V.~V., 1992, Nature, 357, 472 
\bibitem{thompson2007} Thompson T.~A., 2007, RMxAC, 27, 80 
\bibitem{abbott2008} Abbott B., et al., 2008, ApJ, 681, 1419 
\bibitem{abadie2012} Abadie J., et al., 2012, ApJ, 755, 2 
\bibitem{aasi2013} Aasi, J., et al., 2013, arXiv, arXiv:1304.0670 
\bibitem{corsi2009} Corsi A., M{\'e}sz{\'a}ros P., 2009, ApJ, 702, 1171
\bibitem{li1998} Li L.-X., Paczy{\'n}ski B., 1998, ApJ, 507, L59 
\bibitem{kulkarni2005} Kulkarni S.~R., 2005, astro, arXiv:astro-ph/0510256 
\bibitem{metzger2010} Metzger B.~D., et al., 2010, MNRAS, 406, 2650 
\bibitem{piran2013} Piran T., Nakar E., Rosswog S., 2013, MNRAS, 430, 2121 
\bibitem{zhang2013} Zhang B., 2013, ApJ, 763, L22 
\bibitem{gao2013} Gao H., Ding X., Wu X.-F., Zhang B., Dai Z.-G., 2013, ApJ, 771, 86 

\end{thebibliography}
